\documentclass[12pt,a4paper]{article}
\usepackage[T1]{fontenc}
\usepackage[dvipdfmx]{graphicx}
\usepackage{bm,caption,latexsym,amssymb,amsfonts,mathtools,here,ulem}
\usepackage{bm,latexsym,amsfonts,mathtools}
\usepackage{amssymb, amsmath, comment}
\usepackage{color}
\usepackage{hyperref}
\usepackage{booktabs}
\usepackage{mathrsfs}
\usepackage{wrapfig}
\usepackage{setspace}
\numberwithin{equation}{section}
\renewcommand{\baselinestretch}{1.5}

\usepackage[sorting=none,doi=false]{biblatex}

\bibliography{reference}

\begin{document}
\begin{titlepage}
\unitlength = 1mm
\begin{flushright}
KOBE-COSMO-23-01
\end{flushright}

\vskip 1cm
\begin{center}

{ \large \textbf{
A peak in the power spectrum of primordial gravitational waves induced by primordial dark magnetic fields
}}
\vspace{1.8cm}
\\
Sugumi Kanno$^*$, Ann Mukuno$^{\flat}$, Jiro Soda$^{\flat,\sharp}$, and Kazushige Ueda$^{*,\flat}$
\vspace{1cm}

\shortstack[l]
{\it $^*$ Department of Physics, Kyushu University, 744 Motooka, Nishi-ku, Fukuoka 819-0395, Japan \\ 
\it $^\flat$ Department of Physics, Kobe University, Kobe 657-8501, Japan\\
\it $^\sharp$ International Center for Quantum-field Measurement Systems for Studies
of the Universe and\\ 
\ \  \it  Particles (QUP), KEK, Tsukuba 305-0801, Japan
}

\vskip 4.0cm

{\large Abstract}\\
\end{center}
Dark gauge fields have been discussed as  candidates for dark matter recently. If they existed, primordial dark magnetic fields during inflation would have existed. It is believed that primordial gravitational waves (PGWs) arise out of quantum fluctuations during inflation. We study the graviton-dark photon conversion process in the presence of background primordial dark magnetic fields and find that the process induces the tachyonic instability of the PGWs. As a consequence, a peak appears in the power spectrum of PGWs. It turns out that the peak height depends on the direction of observation. The peak frequency could be in the range from $10^{-5}$ to $10^{3}$ Hertz for GUT scale inflation. Hence, the observation of PGWs could provide a new window for probing primordial dark magnetic fields.

\vspace{1.0cm}
\end{titlepage}

\hrule height 0.075mm depth 0.075mm width 165mm
\tableofcontents
\vspace{1.0cm}
\hrule height 0.075mm depth 0.075mm width 165mm
\section{Introduction}

Since the first direct detection of gravitational waves~\cite{LIGOScientific:2016aoc}, gravitational waves have been the most important tool for investigating fundamental physics. Gravitational waves interact with matter very weakly and travel through the universe virtually unimpeded, so they can be a probe of the early inflationary universe. The inflationary scenario predicts that primordial gravitational waves (PGWs) are generated from quantum fluctuations of spacetime. In terms of the density parameter,
the spectrum $\Omega_{\rm GW} (f)$ for the GUT scale inflation is independent of frequency $f$, that is, of wave number $k$ for the gravitons created at the de Sitter to radiation dominant transition:
\begin{eqnarray}
    \Omega_{\rm GW} (f)
    = 10^{-14} \left(\frac{H}{10^{-4} M_{\rm pl}}\right)^2 \ ,
    \label{1.1}
\end{eqnarray}
where $H$ is the Hubble parameter, $M_{\rm pl}$ is the reduced Planck mass. This scale-invariant power spectrum comes from the time translation invariance of de Sitter space.
There are several experimental projects for detecting PGWs~\cite{Hazumi:2019lys,LiteBIRD:2022cnt,Kawamura:2011zz, Amaro-Seoane:2012aqc}. 
If the nonclassicality of the PGWs is found, it implies the existence of gravitons. 
Notably, due to the particle creation, the quantum state of gravitons becomes squeezed during inflation  ~\cite{Grishchuk:1989ss,Grishchuk:1990bj,Albrecht:1992kf,Polarski:1995jg,Kanno:2021vwu}. 
Since the squeezing tends to enhance the observability of gravitons, it is expected that we may be able to probe quantum gravity through observations of PGWs. There are new ideas for detecting the quantum nature of PGWs. One method is to utilize the Hanbury Brown-Twiss interferometry developed in quantum optics for the PGWs, which can distinguish nonclassical particles from classical ones by measuring intensity-intensity correlations~\cite{Giovannini:2010xg,Kanno:2018cuk,Kanno:2019gqw}.  Moreover,  the squeezed state of gravitons can be measured indirectly through their noise in the interferometers~\cite{Parikh:2020nrd,Kanno:2020usf,Parikh:2020kfh,Parikh:2020fhy} or by measuring the decoherence time of a quantum object caused by the surrounding primordial gravitons~\cite{Kanno:2021gpt}.

In the presence of primordial magnetic fields, the effect of the magnetic fields on PGWs is studied in~\cite{Kanno:2022ykw}. In the case of scale-invariant electromagnetic fields, there arises an entanglement between gravitons and photons~\cite{Kanno:2022kve}.
Recently, a possibility of dark photons has been intensively studied. Once the dark photon is allowed, it is legitimate to assume the existence of primordial dark magnetic fields during inflation~\cite{Masaki:2018eut}. 
Therefore, the presence of scale-invariant dark electromagnetic fields may affect the squeezed state of gravitons. 
Thus, it is important to clarify the effect of primordial dark magnetic fields on PGWs.

In this paper, we consider the background primordial dark magnetic fields during inflation and the scale-invariant perturbed dark electromagnetic fields. 
It is known that the presence of background  magnetic fields causes the conversion of gravitons into photons and vice versa~\cite{Gertsenshtein:1962,Raffelt:1987im}. We focus on the effect of the graviton-dark photon conversion process on the PGW power spectrum.
We show that there occurs the tachyonic instability of PGWs due to graviton-dark photon conversion in the presence of primordial dark magnetic fields. Because of this instability, 
there arises a peak in the PGW spectrum (\ref{1.1}). Moreover, since the primordial dark magnetic fields specify a direction in the space, there appears statistical anisotropy in the power spectrum of PGWs. Remarkably, it turns out that  the peak frequency could be in the range from $10^{-5}$ to $10^{3}$ Hertz. 
The presence of the peak turns out to enhance the squeezing of gravitons and makes it easy to observe the PGWs. More interestingly, we may be able to probe the primordial dark magnetic fields by observing the spectrum of PGWs. 

The organization of the paper is as follows. In section 2, 
we present the setup. In section 3, we calculate 
Bogoliubov coefficients. In section 4, we calculate the PGW power spectrum.
We see the tachyonic instability leads to a peak in the spectrum. We also discuss the observability of the peak. 
Section 5 is devoted to the conclusion.

\section{Graviton-dark photon interaction in the early universe }

We consider the Einstein-Hilbert action and the action for a $U(1)$ gauge field coupled with a scalar field:
\begin{eqnarray}
S=S_g+S_\phi+S_A
=\int d^4x \sqrt{-g}\,
\left[
\frac{M_{\rm pl}^2}{2}
\,R
-\frac{1}{2}(\partial_\mu \phi)(\partial^\mu \phi)-V(\phi)
-\frac{1}{4}
f^2(\phi)
F^{\mu\nu}
F_{\mu\nu}
\right]
\label{original action}\,,
\end{eqnarray}
where $M_{\rm pl}=1/\sqrt{8\pi G}$ is the Planck mass. 
The dark gauge field $A_\mu$ represents dark photons and the field strength is defined by $F_{\mu\nu}=\partial_\mu A_{\nu}-\partial_\nu A_{\mu}$.

The background inflationary dynamics is determined
by the metric 
\begin{eqnarray}
ds^2=a^2(\eta)\left[-d\eta^2+\delta_{ij} dx^idx^j\right]\,,
\end{eqnarray}
 and the inflaton $\phi (\eta)$. Once the background is given,
 the coupling function can be regarded as a function of the conformal time $\eta$;
$
f =  f(\eta) \ .
$
We also assume the presence of constant dark magnetic fields
$
 B_i =  {\rm constant} \,.
$
It should be emphasized that the physical dark magnetic fields
are not $B_i$ but $f B_i$. 
In the following, we consider the quantum evolution of 
gravitons and dark 
 photons in the above background.

\subsection{Primordial GWs}
We consider gravitons in a spatially flat expanding background represented by tensor mode perturbations in the three-dimensional metric $h_{ij}$, 
\begin{eqnarray}
ds^2=a^2(\eta)\left[-d\eta^2+\left(\delta_{ij}+h_{ij}\right)dx^idx^j\right]\,,
\end{eqnarray}
where $h_{ij}$ satisfies the transverse traceless conditions $h_{ij}{}^{,j}=h^i{}_i=0$. The spatial indices $i,j,k,\cdots$ are raised and lowered by $\delta^{ij}$ and $\delta_{k\ell}$. In the case of de Sitter space, the scale factor is given by $a(\eta)=-1/(H\eta)$ where $-\infty<\eta<0$.

Expanding the Einstein-Hilbert action up to the second order in perturbations $h_{ij}$, we have
\begin{eqnarray}
\delta S_g=\frac{M_{\rm pl}^2}{8}\int d^4x\,a^2\left[
h^{ij\prime}\,h_{ij}^\prime-h^{ij,k}h_{ij,k}
\right] \,,
\label{action:g}
\end{eqnarray}
where a prime denotes the derivative with respect to the conformal time.
At this quadratic order of the action, it is convenient to expand $h_{ij}(\eta,x^i)$  in Fourier modes,
\begin{eqnarray}
h_{ij}(\eta,x^i)=\frac{2}{M_{\rm pl}} \sum_{P}\frac{1}{(2\pi)^{3/2}} \int d^3 k\,h^{P}_{\bm k}(\eta)\, e_{ij}^{P}(\bm{k})\,e^{i\bm{k}\cdot\bm{x}} 
\ ,
\label{fourier_h}
\end{eqnarray}
where three-vectors are denoted by bold math type and  $e_{ij}^{P}(\bm{k})$  are the polarization tensors  for the ${\bm k}$ mode  normalized as $e^{ijP}(\bm{k})e_{ij}^{Q}(\bm{k})=\delta^{PQ}$ with $P,Q=+,\times$. Then the action (\ref{action:g}) in the Fourier space becomes
\begin{eqnarray}
\delta S_g=\frac{1}{2}\sum_{P}\int d^3k\,d\eta\,a^2\left[\,
|h_{\bm k}^{P\prime}|^2-k^2|h_{\bm k}^P|^2
\,\right]\,.
\label{action_fourier_h}
\end{eqnarray}

\subsection{Primordial dark magnetic fields}

Next, we consider the action for the dark photon up to the second order in perturbations $A_i$, which is given by
\begin{eqnarray}
\delta S_A=\frac{1}{2}\int d^4x\ f^2 \left[A_i^{\prime\, 2}-A_{k,i}^2\right]\,,
\label{action:A}
\end{eqnarray}
where the dark photon field satisfies the Coulomb gauge $A_0=0$ and $A^i{}_{,i}=0$.

Expanding the field $A_i(\eta,x^i)$ by the Fourier modes, we find
\begin{align}
A_i(\eta,x^i)=\sum_{P} \frac{\pm i}{(2\pi)^{3/2}}
\int d^3 k\,A^{P}_{\bm k}(\eta)\,e_i^{P}(\bm{k})\, e^{i\bm{k}\cdot\bm{x}}
\label{fourier_A}
\ ,
\end{align}
where  $e_i^{P}(\bm{k})$ are the polarization  vectors for the ${\bm k}$ mode  normalized as $e^{iP}(\bm{k}) e_i^{Q}(\bm{k})=\delta^{PQ}$ with $P,Q=+,\times$. The sign of $\pm i$ corresponds to the $P,Q=+,\times$.
The action (\ref{action:A}) in terms of the Fourier modes reads
\begin{eqnarray}
\delta S_A=\frac{1}{2}\sum_{P}\int d^3k\,d\eta\,f^2\left[\,
|A_{\bm k}^{P\prime}|^2-k^2|A_{\bm k}^P|^2
\,\right]\,.
\label{action_fourier_A}
\end{eqnarray}

\subsection{Graviton-dark photon conversion}
The action for the interaction between the graviton and the dark photon  up to second order in perturbations $h_{ij}, A^i$ is found to be
\begin{eqnarray}
\delta S_{\rm I}=\int d^4x \left[
\varepsilon_{i\ell m} f^2 B_m h^{ij}\left(\partial_j A_\ell
-\partial_\ell A_j\right)
\right]\,,
\label{action:I}
\end{eqnarray}
where $B_m=\varepsilon_{mj\ell}\,\partial_j A_\ell$ is a constant background dark magnetic field.

\begin{figure}[H]
\centering
 \includegraphics[keepaspectratio, scale=0.55]{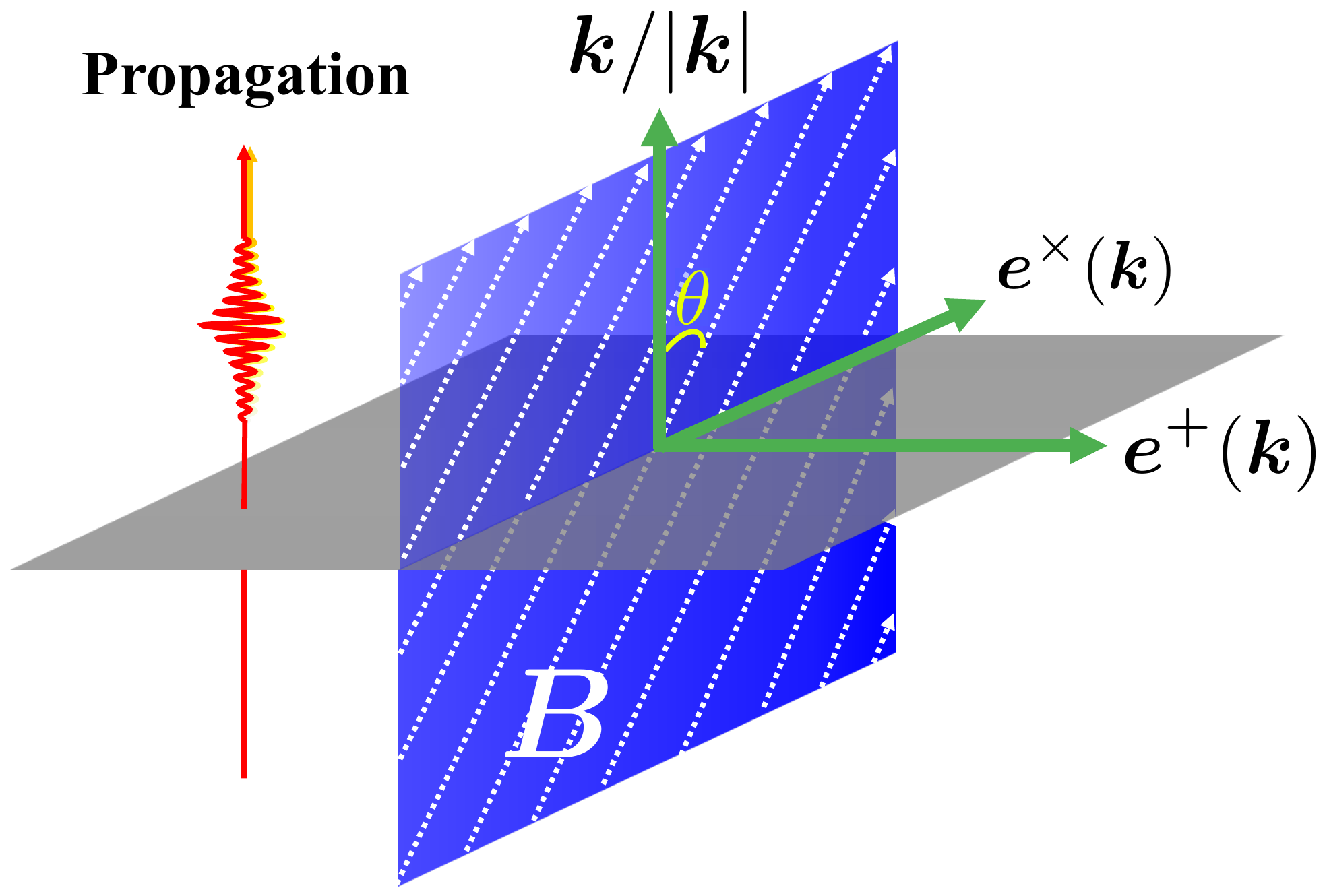} \renewcommand{\baselinestretch}{3}
 \caption{Configuration of the polarization vector ${\bm e}^P(\bm k)$, wave number ${\bm k}$, and background dark magnetic field ${\bm B}$. The angle between the dark magnetic field and the wavenumber vector is represented by $\theta$.}
 \label{Configuration}
 \end{figure}

In terms of the Fourier modes defined in Eqs.~(\ref{fourier_h}) and (\ref{fourier_A}), 
the interaction term reads 
\begin{eqnarray}
\delta S_I = \frac{2}{M_{\rm pl}}\sum_{P,Q}\int d^3 k\,d\eta\,f^2\left[
\varepsilon_{i\ell m}\,B_m\,h_{\bm k}^PA_{-\bm k}^Q
\,e_{ij}^P(\bm k)\Bigl\{ik_\ell\,e_{j}^Q(-\bm k)-ik_j\,e_{\ell }^Q(-\bm k)\Bigr\}\right]
\label{action:I2}\,,
\end{eqnarray}
where $k=|\bm k|$. Polarization vectors $e^{i+}, e^{i\times}$ and a vector $k^i/k$ constitute an orthonormal basis.
Without loss of generality, we assume the constant background dark magnetic field is in the ($k^i, e^{i \times}$)-plane as depicted in FIG.~\ref{Configuration}.
\noindent
The polarization tensors  can be written in terms of polarization vectors  
$e^{i+}$ and $e^{i\times}$
 as
\begin{align}
    &e_{ij}^+(\bm{k})
    =\frac{1}{\sqrt{2}} \Bigl\{
    e^+_i(\bm{k}) e^+_j(\bm{k})-e^\times_i(\bm{k}) e^\times_j(\bm{k})
    \Bigr\}\,,\\
    &e_{ij}^\times(\bm{k})
    =\frac{1}{\sqrt{2}} 
    \Bigl\{
    e^+_i(\bm{k}) e^\times_j(\bm{k})+e^\times_i(\bm{k}) e^+_j(\bm{k})
    \Bigr\}\, .
\end{align}
Below, we assume
$
e_i^\times(-\bm{k})=-e_i^\times(\bm{k})
$.
The action (\ref{action:I2}) is then written as
\begin{eqnarray}
\delta S_I&=&\int d^3k\,d\eta\,f^2\,
\lambda\,k \left[\,h_{\bm k}^+(\eta)\,A_{-\bm k}^+(\eta)+\,h_{\bm k}^\times(\eta)\,A_{-\bm k}^\times(\eta)\,\right]\, ,
\label{action:I3}
\end{eqnarray}
where we defined the coupling between graviton and dark photon as 
\begin{align}
\lambda
\equiv
\frac{\sqrt{2}}{M_{\rm pl}}
\varepsilon^{i\ell m}\,e_i^+\,\frac{k_\ell}{k}\,B_m\,.
\label{coupling}
\end{align}
Here, the conditions for the graviton and dark photon to be real are read,
$h_{-\bm k}^{+,\times}(\eta)=h_{\bm k}^{*\,+,\times}(\eta)$ and $A_{-\bm k}^{+,\times}(\eta)=-A_{\bm k}^{*\,+,\times}(\eta)$\,.
Below, we focus on the plus polarization and omit the index $P$ unless there may be any confusion. 

\subsection{Diagonal equations}

If we use the canonical variable $y_{\bm k}(\eta)=a\,h_{\bm k}(\eta)$ and $x_{\bm k}(\eta)=f\,A_{\bm k}(\eta)$, the total action of Eqs.~(\ref{action_fourier_h}), (\ref{action_fourier_A}) and (\ref{action:I3}) are written as
\begin{eqnarray}
\delta S&=&\delta S_y+\delta S_x+\delta S_I
\nonumber\\
&=&\frac{1}{2}\int d^3 k\,d\eta\left[\,
|y_{\bm k}^{\prime}|^2
-\left(k^2-\left(\frac{a^\prime}{a}\right)^2\right)|y_{\bm k}|^2
-\frac{a^\prime}{a}\left(
y_{\bm k}\,y_{-\bm k}^{\prime}
+y_{-\bm k}\,y_{\bm k}^{\prime}
\right)
\right]\nonumber\\
&&+\frac{1}{2}\int d^3 k\,d\eta\left[\,
|x_{\bm k}^{\prime}|^2
-\left(k^2-\left(\frac{f^\prime}{f}\right)^2\right)|x_{\bm k}|^2
-\frac{f^\prime}{f}\left(
x_{\bm k}\,x_{-\bm k}^{\prime}
+x_{-\bm k}\,x_{\bm k}^{\prime}
\right)
\right]\,\nonumber\\
&&+\int d^3 k\,d\eta\left[\,\frac{f}{a}\,\lambda\,k\,
y_{\bm k}\,x_{-\bm k}
\,\right]\,.
\label{totalaction}  
\end{eqnarray}
The variation of the action (\ref{totalaction}) with respect to the graviton and the dark photon fields gives
\begin{eqnarray}
    y_{\bm{k}}''+\left(k^2-\frac{a^{\prime\prime}}{a}\right)y_{\bm{k}}=-\lambda\,k\,f\,\frac{x_{\bm k}}{a} \ ,
    \qquad
    x_{\bm{k}}''+\left(k^2-\frac{f^{\prime\prime}}{f}\right)x_{\bm{k}}=-\lambda\,k\,f\,\frac{y_{\bm k}}{a} \ .
    \label{systemofeom}
\end{eqnarray}

We assume the gauge kinetic function in the form 
\begin{eqnarray}
  f(\eta)
  =a(\eta)^{-2c},
  \label{f}
\end{eqnarray}
where $c$ is a constant parameter.
We take $c=-1/2$ to make the analysis easier\footnote{
Note that the gauge field we consider is a dark gauge field, so there appears no strong coupling problem. However, 
in the case of the gauge field, there occur strong coupling~\cite{Demozzi:2009fu,BazrafshanMoghaddam:2017zgx} 
 and back-reaction problems~\cite{Demozzi:2009fu,Kanno:2009ei} when $c=-1/2$ that is, the gauge kinetic function $f$ is a growing function.
However, the gauge kinetic function can be modified so that the strong coupling disappears~\cite{Ferreira:2014hma}.
Even if the strong coupling affects the electric fields, there may be a model where the magnetic fields remain the same. 
}. For this parameter, the power spectrum of the dark electromagnetic fields $A_\mu$ is scale-invariant~\cite{Kanno:2009ei}. While the spectrum of primordial dark magnetic fields becomes $P(B) \propto k^2$.
 Then Eqs.~(\ref{systemofeom}) become
\begin{eqnarray}
    y_{\bm{k}}''+\left(k^2-\frac{a^{\prime\prime}}{a}\right)y_{\bm{k}}=-\lambda\,k\,x_{\bm{k}} \ ,
    \qquad
    x_{\bm{k}}''+\left(k^2-\frac{a^{\prime\prime}}{a}\right)x_{\bm{k}}=-\lambda\,k\,y_{\bm{k}} \ .
    \label{systemofeom2}
\end{eqnarray}
If we define new variables $X_{\bm{k}}$ and $Y_{\bm{k}}$ such as
\begin{eqnarray}
Y_{\bm{k}}=\frac{1}{\sqrt{2}}\left(y_{\bm{k}}+x_{\bm{k}}\right)\, , \qquad
X_{\bm{k}}=\frac{1}{\sqrt{2}}\left(y_{\bm{k}}-x_{\bm{k}}\right)\, .
\end{eqnarray}
Eq.~(\ref{systemofeom2}) are diagonalized in the form
\begin{eqnarray}
Y_{\bm{k}}^{\prime\prime}+\left(k^2+\lambda\,k-\frac{a^{\prime\prime}}{a}\,\right)Y_{\bm{k}}=0\,\ ,
\qquad
X_{\bm{k}}^{\prime\prime}+\left(k^2-\lambda\,k-\frac{a^{\prime\prime}}{a}\,\right)X_{\bm{k}}=0 \ .
\label{eom:Y}
\end{eqnarray}

\section{Generation and evolution of primordial GWs}

In this section, we present a cosmological setup in order to calculate Bogoliubov coefficients. 

\subsection{Cosmological setup}

We consider instantaneous reheating after inflation approximated by the de Sitter phase leading to a radiation-dominated phase. The scale factor evolves as follows
\begin{eqnarray}
 a (\eta )=\left\{
\begin{array}{l}
\vspace{0.2cm}
 \frac{1}{-H  (  \eta -2 \eta_1 )}  \hspace{1.2cm}   {\rm for}  \quad   \eta   <  \eta_1\,,\\
\frac{\eta}{H\eta^2_1} \hspace{2.2cm}     {\rm for} \quad     0<\eta_1  <  \eta   \ .
\end{array}
\right.             
\label{history}
\end{eqnarray}
where $H$ is the Hubble parameter during inflation and $\eta_1$ is the time of reheating. The scale factor is smoothly connected up to the first order of derivative at the reheating. 
We assume that dark magnetic fields arise instantaneously from the Bunch-Davies vacuum at the time $\eta_*$.
When $c<-1$, the dark magnetic fields are
 generated dynamically during inflation. The situation is quite similar to anisotropic inflation~\cite{Watanabe:2009ct}.

\begin{figure}[H]
\centering
 \includegraphics[scale=0.55]{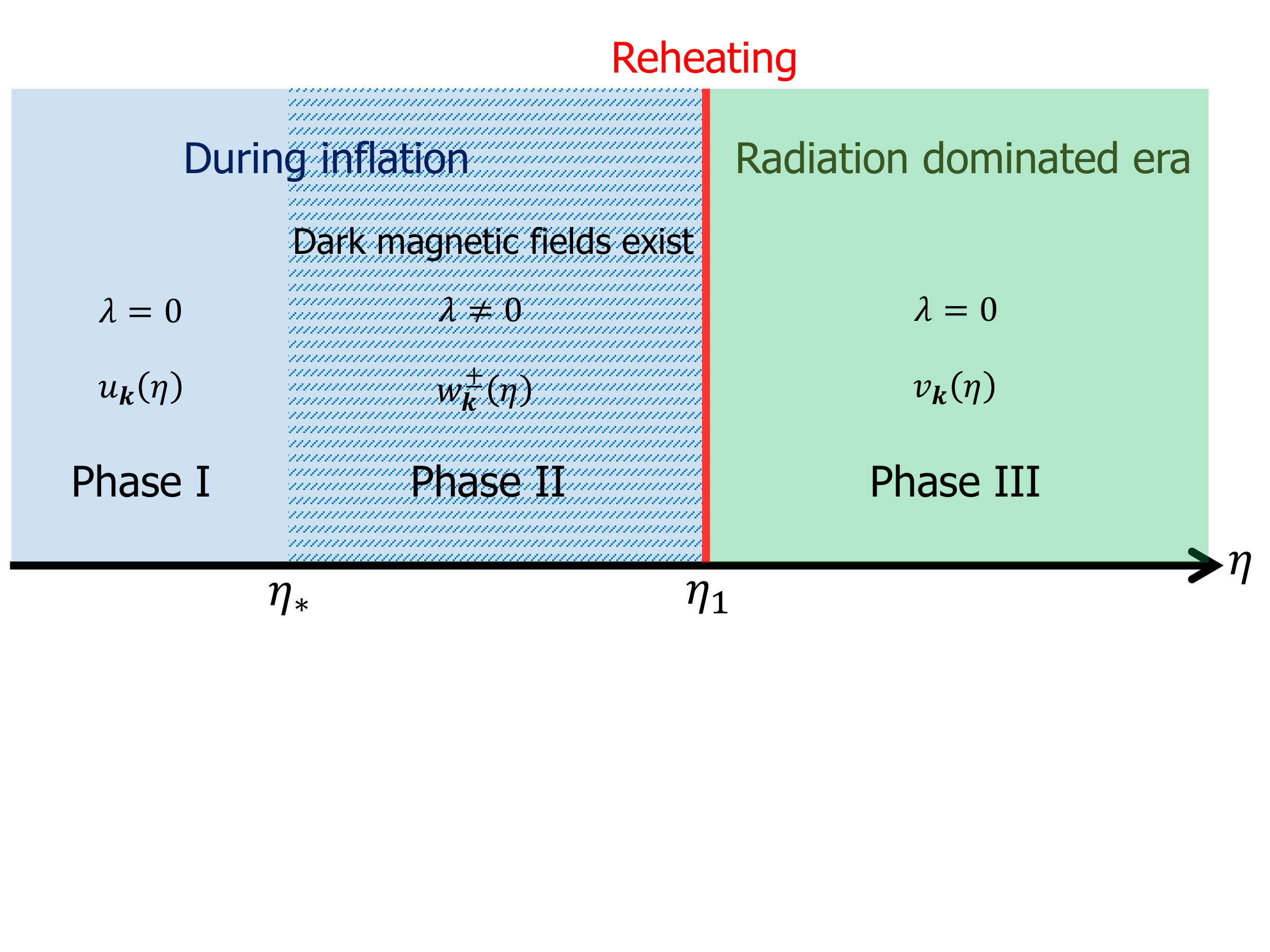} \renewcommand{\baselinestretch}{3}
 \vspace{-3.5cm}
 \caption{During inflation, dark magnetic fields $B_i$ arise instantaneously at $\eta_*$. Instantaneous reheating at $\eta_1$ leads to a radiation-dominated phase.}
 \label{GWPS1}
 \end{figure}
\noindent
Before the time $\eta_*$ (phase I), the equations of motion read
\begin{eqnarray}
    Y_{\bm{k}}''+\left(k^2-\frac{2}{(\eta-2\eta_1)^2}\right)Y_{\bm{k}}=0 \ ,
    \qquad
    X_{\bm{k}}''+\left(k^2-\frac{2}{(\eta-2\eta_1)^2}\right)X_{\bm{k}}=0 
    \label{eom:I}\ .
\end{eqnarray}
After the $\eta_*$, the coupling $\lambda$ is switched on until $\eta_1$ (phase II). There, gravitons and dark photons are coupled as
\begin{eqnarray}
    Y_{\bm{k}}''+\left(k^2+\lambda\,k-\frac{2}{(\eta-2\eta_1)^2}\right)Y_{\bm{k}}=0 \ ,
    \qquad
    X_{\bm{k}}''+\left(k^2-\lambda\,k-\frac{2}{(\eta-2\eta_1)^2}\right)X_{\bm{k}}=0 \ 
    \label{eom:II}.
\end{eqnarray}
After the reheating time $\eta_1$ (phase III), 
we assume $f={\rm const}.$ Then, the equations of motion become
\begin{eqnarray}
    Y_{\bm{k}}''+k^2\,Y_{\bm{k}}=0 \ ,
    \qquad
    X_{\bm{k}}''+k^2\,X_{\bm{k}}=0 \ 
    \label{eom:III}.
\end{eqnarray}
We assume the vacuum state can be specified by the Bunch-Davies vacuum which behaves in the same way as the Minkowski vacuum in the remote past in the de Sitter phase (Phase I). In the Phase II, the basis is properly normalized but deformed due to the presence of the dark magnetic field. In the Phase III, we choose the positive frequency mode for the adiabatic vacuum as a basis in radiation dominated phase in order to discuss particle creation. Thus we choose the properly normalized basis in each phase as the solution of Eqs.~(\ref{eom:I}), (\ref{eom:II}) and (\ref{eom:III}).
\begin{eqnarray}
u_{\bm{k}} &=& \frac{1}{\sqrt{2k}}\left(1-\frac{i}{k\left(\eta-2\eta_1\right)}\right)e^{-i k\left(\eta-2\eta_1\right)}\hspace{4.4cm}   {\rm for}  \quad {\rm Phase~I}\quad\label{u}\\
\quad w^{+}_{\bm{k}} &=& \frac{1}{\sqrt{2\sqrt{k^2+\lambda k}}}\left(1-\frac{i}{\sqrt{k^2+\lambda k}\ \left(\eta-2\eta_1\right)}\right)e^{-i \sqrt{k^2+\lambda k}\,\left(\eta-2\eta_1\right)}\hspace{0.2cm}   {\rm for}  \quad {\rm Phase~II}\quad\label{w+}\\
w^{-}_{\bm{k}} &=& \frac{1}{\sqrt{2\sqrt{k^2-\lambda k}}}\left(1-\frac{i}{\sqrt{k^2-\lambda k}\ \left(\eta-2\eta_1\right)}\right)e^{-i \sqrt{k^2-\lambda k}\,\left(\eta-2\eta_1\right)}\hspace{0.2cm}   {\rm for}  \quad {\rm Phase~II}\quad\label{w-}\\
v_{\bm{k}} &=& \frac{1}{\sqrt{2k}}e^{-i k\eta}\,
\hspace{8.6cm}   {\rm for}  \quad {\rm Phase~III}\quad\label{v}
\end{eqnarray}
The above mode functions are normalized as 
\begin{eqnarray}
u_{\bm{k}}u_{\bm{k}}^{*\prime}-u_{\bm{k}}^* u_{\bm{k}}^{\prime}=i ,\quad
w^{\pm}_{\bm{k}}w_{\bm{k}}^{\pm*\prime}-w_{\bm{k}}^{\pm*} w_{\bm{k}}^{\pm\prime}=i , \quad
v_{\bm{k}}v_{\bm{k}}^{*\prime}-v_{\bm{k}}^* v_{\bm{k}}^{\prime}=i\,.
\label{normalization}
\end{eqnarray}

\subsection{Bogoliubov Transformations}

Through the history from Phase I to III, the fields $X_{\bm{k}}$ and $Y_{\bm{k}}$ in Eqs.~(\ref{eom:Y}) can be expanded by the mode functions $U_{\bm{k}}$ and $V_{\bm{k}}$, respectively such as
\begin{eqnarray}
Y_{\bm{k}} = \hat{a}_{\bm{k}}U_{\bm{k}}
+\hat{a}_{-\bm{k}}^\dag U_{\bm{k}}^* \ , \qquad
X_{\bm{k}} = \hat{b}_{\bm{k}}V_{\bm{k}}
+\hat{b}_{-\bm{k}}^\dag V_{\bm{k}}^*\,.
\label{expansion}
\end{eqnarray}
The commutation relations $[\hat{a}_{\bm k},\hat{a}^\dag_{\bm k^\prime}]=\delta({\bm k}-{\bm k}^\prime)$ and $[\hat{b}_{\bm k},\hat{b}^\dag_{\bm k^\prime}]=\delta({\bm k}-{\bm k}^\prime)$ guarantee the canonical commutation relations.
We write the mode function $U_{\bm{k}}/V_{\bm{k}}$ that satisfies Eqs.~(\ref{eom:I}), (\ref{eom:II}) and (\ref{eom:III}) as $U_{\bm{k}}^{\rm I}/V_{\bm{k}}^{\rm I}$, $U_{\bm{k}}^{\rm II}/V_{\bm{k}}^{\rm II}$ and $U_{\bm{k}}^{\rm III}/V_{\bm{k}}^{\rm III}$.

As the initial state in Phase I, we consider the Bunch-Davies vacuum $\hat{a}_{\bm{k}}|0\rangle_{\rm BD}=0$, $\hat{b}_{\bm{k}}|0\rangle_{\rm BD}=0$. Then the mode functions $U_{\bm{k}}$ and $V_{\bm{k}}$ are expressed by the positive frequency mode Eq.~(\ref{u}):
\begin{eqnarray}
U_{\bm{k}}^{\rm I} = V_{\bm{k}}^{\rm I} = u_{\bm{k}}(\eta).
\label{UI}
\end{eqnarray}
In phase II, the mode function $U_{\bm{k}}$ and $V_{\bm{k}}$ are written by the solutions Eqs.~(\ref{w+}) and (\ref{w-}) of Eqs.~(\ref{eom:II}) and their complex conjugate,
\begin{eqnarray}
U_{\bm{k}}^{\rm II} &=& \alpha_{{\rm II}k}^{+}\,w^{+}_k (\eta  )  + \beta_{{\rm II}k}^{+}\, w^{+*}_{k} (\eta) \ ,
\label{UII}\\
V_{\bm{k}}^{\rm II} &=& \alpha_{{\rm II}k}^{-}\,w^{-}_k (\eta  )  + \beta_{{\rm II}k}^{-}\, w^{-*}_{k} (\eta)  \,,
\end{eqnarray}
where $\alpha_{{\rm II}k}^{\pm}$ and $\beta_{{\rm II}k}^{\pm}$ are integration constants which can be interpreted as the Bogoliubov coefficients. Similarly, in phase III, the mode functions are written by the solution Eq.~(\ref{v}) of Eq.~(\ref{eom:III}):
\begin{eqnarray}
U_{\bm{k}}^{\rm III} &=&\alpha_{{\rm III}k}^{+}\,v_k (\eta  )  + \beta_{{\rm III}k}^{+}\, v^*_{k} (\eta) \ ,
\label{UIII}\\
V_{\bm{k}}^{\rm III} &=&\alpha_{{\rm III}k}^{-}\,v_k (\eta  )  + \beta_{{\rm III}k}^{-}\, v^*_{k} (\eta)  \ ,
\end{eqnarray}
where $\alpha_{{\rm III}k}^{\pm}$ and $\beta_{{\rm III}k}^{\pm}$ are integration constants interpreted as the Bogoliubov coefficients. 

Since the solutions of $Y_{\bm{k}}$ and $X_{\bm{k}}$ must be continuous and continuously differentiable, 
the boundary conditions for the above mode functions are required at $\eta_*$ and $\eta_1$ respectively. Let us focus on $Y_{\bm{k}}$. The boundary conditions at $\eta_*$ are given by, 
\begin{eqnarray}
&&U_{\bm{k}}^{\rm I}(\eta_*)=U_{\bm{k}}^{\rm II}(\eta_*)\,\\
&&U_{\bm{k}}^{{\rm I}\,\prime}(\eta)\Big|_{\eta=\eta_*}=U_{\bm{k}}^{{\rm II}\,\prime}(\eta)\Big|_{\eta=\eta_*}\,
\end{eqnarray}
The same is true for the $V_{\bm{k}}^{\rm I}(\eta_0)$.
Plugging Eqs.~(\ref{UI}) and (\ref{UII}) into the above, we have
\begin{eqnarray}
\begin{pmatrix}
u(\eta_*)\\
u^\prime(\eta_*)\\
\end{pmatrix}
=
\begin{pmatrix}
w^+_k (\eta_*) & w^{+*}_{k} (\eta_*)\\
w^{+\,\prime}_{k} (\eta_*) & w^{+*\,\prime}_k (\eta_*)\\
\end{pmatrix}
\begin{pmatrix}
\alpha_{{\rm II}k}^+ \\
\beta_{{\rm II}k}^+\\
\end{pmatrix}. \quad 
\label{boundary1}
\end{eqnarray}
Similarly, the boundary conditions at $\eta_1$ are
\begin{eqnarray}
&&U_{\bm{k}}^{\rm II}(\eta_1)=U_{\bm{k}}^{\rm III}(\eta_1)\,\\
&&U_{\bm{k}}^{{\rm II}\,\prime}(\eta)\Big|_{\eta=\eta_1}=U_{\bm{k}}^{{\rm III}\,\prime}(\eta)\Big|_{\eta=\eta_1}\,
\end{eqnarray}
Plugging Eqs.~(\ref{UII}) and (\ref{UIII}) into the above, we have
\begin{eqnarray}
\begin{pmatrix}
w^+_k (\eta_1) & w^{+*}_{k} (\eta_1)\\
w^{+\,\prime}_{k} (\eta_1) & w^{+*\,\prime}_k (\eta_1)\\
\end{pmatrix}
\begin{pmatrix}
\alpha_{{\rm II}k}^+ \\
\beta_{{\rm II}k}^+\\
\end{pmatrix}
=
\begin{pmatrix}
v_k (\eta_1) & v^{*}_{k} (\eta_1)\\
v^{\,\prime}_{k} (\eta_1) & v^{*\,\prime}_k (\eta_1)\\
\end{pmatrix}
\begin{pmatrix}
\alpha_{{\rm III}k}^+ \\
\beta_{{\rm III}k}^+\\
\end{pmatrix}. \quad 
\label{boundary2}
\end{eqnarray}
Combining Eqs.~(\ref{boundary1}) and (\ref{boundary2}), we find
\begin{eqnarray}
\begin{pmatrix}
\alpha_{{\rm III}k}^+ \\
\beta_{{\rm III}k}^+\\
\end{pmatrix}
=
\begin{pmatrix}
v_k (\eta_1) & v^{*}_{k} (\eta_1)\\
v^{\,\prime}_{k} (\eta_1) & v^{*\,\prime}_k (\eta_1)\\
\end{pmatrix}^{\!\!-1}
\begin{pmatrix}
w^+_k (\eta_1) & w^{+*}_{k} (\eta_1)\\
w^{+\,\prime}_{k} (\eta_1) & w^{+*\,\prime}_k (\eta_1)\\
\end{pmatrix} 
\begin{pmatrix}
w^+_k (\eta_*) & w^{+*}_{k} (\eta_*)\\
w^{+\,\prime}_{k} (\eta_*) & w^{+*\,\prime}_k (\eta_*)\\
\end{pmatrix}^{\!\!-1} 
\begin{pmatrix}
u_k (\eta_*)\\
u'_k (\eta_*)\\
\end{pmatrix}. \quad 
\label{bogoliubov1}
\end{eqnarray}
By using the normalization in Eq.~(\ref{normalization}), the above can be written as
\begin{eqnarray}
\begin{pmatrix}
\alpha_{{\rm III}k}^+ \\
\beta_{{\rm III}k}^+\\
\end{pmatrix}
=
\begin{pmatrix}
-v^{*\prime}_{k} (\eta_1) & v^*_{k} (\eta_1)\\
v^{\prime}_{k} (\eta_1) & -v_k (\eta_1  )\\
\end{pmatrix} 
\begin{pmatrix}
w_k^{+} (\eta_1  )  & w^{+*}_{k} (\eta_1)\\
w^{+\prime}_{k} (\eta_1) & w^{+*\prime}_{k}(\eta_1)\\
\end{pmatrix} 
\begin{pmatrix}
w^{+*\prime}_{k} (\eta_*) & -w^{+*}_{k} (\eta_*)\\
-w^{+\prime}_{k} (\eta_*) & w_k^{+} (\eta_*)\\
\end{pmatrix} 
\begin{pmatrix}
u_k (\eta_*)\\
u'_k (\eta_*)\\
\end{pmatrix}. \quad 
\label{bogoliubov2}
\end{eqnarray}
For the $V_{\bm k}$, the result is obtained by replacing the superscript $+$ by $-$.
Using the above results, we calculate the power spectrum
of energy density in the next section.

\section{Tachyonic instability and the PGW power spectrum}

\begin{figure}[H]
\centering
 \includegraphics[keepaspectratio, scale=0.75]{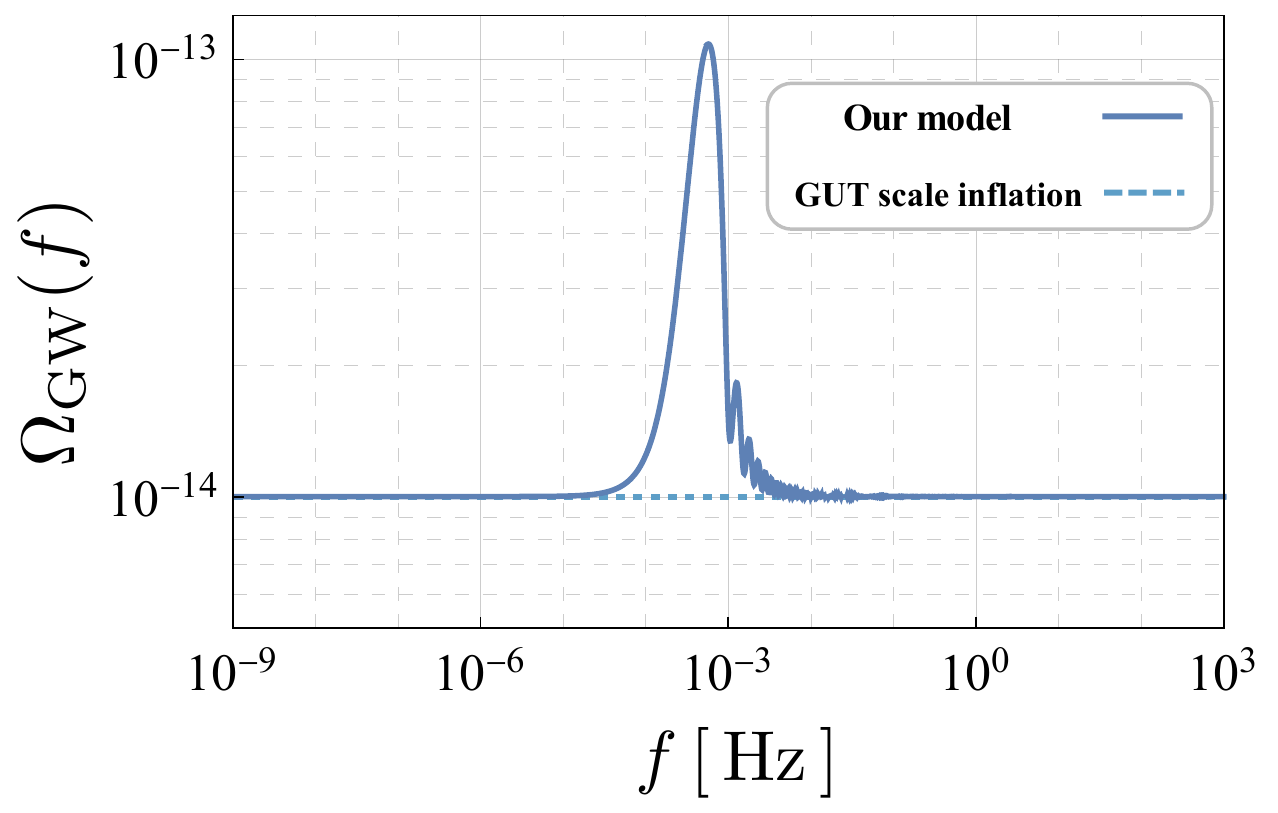} \renewcommand{\baselinestretch}{3}
\caption{GW power spectrum is plotted for $H=10^{14}\, {\rm GeV}$ and $B_0=2\sqrt{3}\times 10^{-16}\,$G. Parameters are set as $\eta_* =-1\,{\rm GeV}^{-1}$, $\eta_1=10^{-14}\,{\rm GeV}^{-1}$, and $\lambda=2\sqrt{3}\,{\rm GeV}$. The GUT scale inflation predicts a flat spectrum of the amplitude ${\Omega}_{\rm GW} = 10^{-14}$.}
 \label{GWPS2}
 \end{figure}

Let us discuss the power spectrum of the energy density of gravitons. From Eq.~(\ref{systemofeom2}), the graviton is given by
\begin{eqnarray}
y_{\bm{k}}=\frac{1}{\sqrt{2}}\left(Y_{\bm{k}}+X_{\bm{k}}\right)\ .
\end{eqnarray}
By using Eq.~(\ref{expansion}), the graviton operator in radiation dominated phase is written as
\begin{eqnarray}
y_{\bm{k}}&=&\frac{1}{\sqrt{2}}\left(
\hat{a}_{\bm{k}}\,U_{\bm{k}}^{\rm III}
+\hat{a}_{-\bm{k}}^\dag U_{\bm{k}}^{{\rm III}*}
+ \hat{b}_{\bm{k}}V_{\bm{k}}^{\rm III}
+\hat{b}_{-\bm{k}}^\dag V_{\bm{k}}^{{\rm III}*} \right)\nonumber \\
&=& \frac{1}{\sqrt{2}} \left[ 
\left(
\alpha_{{\rm III}k}^{+}\,a_{\bm{k}}+\beta_{{\rm III}k}^{+*}a_{\bm{k}}^\dag 
+\alpha_{{\rm III}k}^{-}b_{\bm{k}} 
+\beta_{{\rm III}k}^{-*}b_{\bm{k}}^\dag
\right)v_k + \left(
\beta_{{\rm III}k}^{+}a_{\bm{k}}+\alpha_{{\rm III}k}^{+*}a_{\bm{k}}^\dag
+\beta_{{\rm III}k}^{-}b_{\bm{k}} 
+\alpha_{{\rm III}k}^{-*}b_{\bm{k}}^\dag
\right)v_k^*
\right]\nonumber \\
&\equiv& A_{k} v_k + A_{k}^\dag v_k^*   \ ,
\end{eqnarray}
where we defined natural annihilation and creation operators $A_k, A_k^\dag$ in the radiation-dominated phase.
Thus, we can identify the number operator of gravitons as
\begin{eqnarray}
{}_{\rm BD}\langle0|\,A_{k}^\dag A_{k}\,|0\rangle_{\rm BD}
= \frac{1}{2}\left(\, |\beta_{{\rm III}k}^+|^2 + |\beta_{{\rm III}k}^-|^2\,\right)  \ .
\end{eqnarray}
Then, the GW spectral density parameter can be calculated as
\begin{eqnarray}
  \Omega_{\rm GW}(f) = \frac{k^4}{2\pi^2\rho_c} \left(\,|\beta_{{\rm III}k}^{+}|^2+ |\beta_{{\rm III}k}^{-}|^2\,\right) \ ,
  \label{omegagw}
\end{eqnarray}
where $\rho_c$ is the critical energy density of the universe and $f=k/2\pi$ is a frequency of GWs. The GW spectral density parameter is plotted in FIG.~\ref{GWPS2} where we considered the GUT scale inflation $H=10^{14}\ {\rm GeV}$ and $B_0=2\sqrt{3}\times 10^{-16}$\,G, $\eta_*=-1\,{\rm GeV}^{-1}$.
We see a peak appears at a particular frequency and no more scale-invariant spectrum as in Eq.~(\ref{1.1}).

\noindent

The origin of the peak comes from the equation for the $X_{\bm{k}}$ in Eq.~(\ref{eom:II}). If we rewrite it in the form
\begin{eqnarray}
X_{\bm{k}}''+\left(\left(k-\frac{\lambda}{2}\right)^2-\frac{\lambda^2}{4} -\frac{2}{(\eta-2\eta_1)^2}\right)X_{\bm{k}}=0 \,,
\end{eqnarray}
we see that the contribution from the coupling $\lambda$ causes a tachyonic instability around $k\simeq \lambda/2$.
This kind of instability of the gravitational waves occurs when the off-diagonal part in equations of motion exists. 
From Eq.~(\ref{coupling}), the strength of the coupling constant can be estimated as
\begin{eqnarray}
    \lambda
    = \frac{\sqrt{2}\,|{\bm B}|}{M_{\rm pl}} \sin \theta \ ,
    \label{lambda}
\end{eqnarray}
where $\theta$ is the angle between the direction of the constant background dark magnetic field and the wavenumber vector of GWs. We assume $\sin\theta >0$ below for simplicity. In the case of $\sin\theta <0$, we just switch the role of $X_k$ to $Y_k$. 
From observations, we know the magnitude of current extragalactic magnetic fields 
takes from $10^{-17}\,{\rm Gauss}$ to $10^{-9}$ Gauss.
We assume the primordial dark magnetic fields also lies in the range $10^{-17}\,{\rm G}<|{\bm B}_0| <10^{-9}\,{\rm G}$. 
After inflation, the energy density of radiation decays as $\rho_\gamma\propto a^{-4}$. By combining the Stefan-Boltzman law $\rho_\gamma=\sigma T^4$, we find that the temperature of radiation decays as the Universe expands such as $T\propto 1/a$. Then the ratio of the present scale factor $a_0$ to that of reheating time $a_1$ is found to be  $a_0/a_1=T_{\rm reh}/T_0$. 
Here,  $T_{\rm reh}$ is the reheating temperature. In our model, the physical dark magnetic fields decay as $B_{\rm phy}\propto 1/a$ during inflation and then decay as $B_{\rm phy}\propto 1/a^2$ after the reheating. Hence, the ratio of the dark magnetic field at the reheating time $B_{\rm reh}$ to that at present $B_0$ becomes 
$B_{\rm reh}/B_0=\left(a_0/a_1\right)^2=\left(T_{\rm reh}/T_0\right)^2\sim 10^{54}\,(T_{\rm reh}/ 
10^{-4} M_{\rm pl})^2$ where we used $T_0\sim 2.7\,{\rm K}\sim 10^{-4}$ eV. 
Accordingly, the magnitude of primordial dark magnetic fields at the end of the inflation is found to be 
 \begin{eqnarray}
 10^{37} \left(\frac{T_{\rm reh}}{10^{-4} M_{\rm pl}}\right)^2 {\rm G}< B_{\rm reh} <10^{45} \left(\frac{T_{\rm reh}}{10^{-4} M_{\rm pl}}\right)^2{\rm G} \ ,
 \end{eqnarray}
where $B_{\rm reh}=|{\bm B}|/a_1=|{\bm B}|$ because we used $a_1=1$ as a convention. By using Eq.~(\ref{lambda}), $1\,{\rm G}\sim 10^{-2}\,{\rm eV}^2$ and $M_{\rm pl}\sim 10^{18}$ GeV, the above range   gives the range of maximum coupling $\lambda_{\rm max}=\sqrt{2}B_{\rm reh}/M_{\rm pl}$ 
at $\theta=\pi/2$ such as
\begin{eqnarray}
10^{-1}\left(\frac{T_{\rm reh}}{10^{-4} M_{\rm pl}}\right)^2{\rm GeV}<\lambda_{\rm max}<10^{7}\left(\frac{T_{\rm reh}}{10^{-4} M_{\rm pl}}\right)^2{\rm GeV}\,.
\end{eqnarray}
The wavenumber at the peak of tachyonic instability  $k\simeq\lambda/2$ is translated into the peak frequency observed today and the above lowest value is translated into
\begin{eqnarray}
    f_0 =\frac{\lambda}{4\pi}\frac{a_1}{a_0}
    = 10^{-5}  \sin \theta\left(\frac{B_0}{10^{-17}{\rm G}}\right)
    \left(\frac{T_{\rm reh}}{10^{-4} M_{\rm pl}} \right)  
    \ {\rm Hz} \,,
\end{eqnarray}
where we used $1$ eV $=10^{15}$ Hz. 
Note that the peak frequency depends on the direction of observation. 

Note also that the wavenumber at the peak would be slowly changed as $k\sim \lambda f/2a$ when the value of the $c$ is  slightly different.

Thus, in terms of maximum frequency at $\theta=\pi/2$, the  range $10^{-17}\,{\rm G}<B_0 <10^{-9}\,{\rm G}$ corresponds to the frequency range
\begin{eqnarray}
10^{-5}\left(\frac{T_{\rm reh}}{10^{-4} M_{\rm pl}} \right){\rm Hz}
<f_{0,{\rm max}} <10^{3}\left(\frac{T_{\rm reh}}{10^{-4} M_{\rm pl}} \right){\rm Hz}\,.
\end{eqnarray} 
Without the primordial dark magnetic field, the density parameter of PGWs is known to be $\Omega_{\rm GW}=10^{-14}$ for the GUT scale inflation as in Eq.~(\ref{1.1}). In the presence of the primordial dark magnetic field, Eq.~(\ref{omegagw}) tells us that the peak height of the $\Omega_{\rm GW}$ is given by 
\begin{eqnarray}
    \Omega_{\rm GW} (f_{\rm peak})
    = 10^{-14} \left(\frac{H}{10^{-4} M_{\rm pl}}\right)^2
\frac{1}{2}\exp\left[- \lambda\,\eta_* \right]\,.
\end{eqnarray}
We find that the PGWs power spectrum depends on the direction of observation due to the dependence of $\theta$ in $\lambda$. The peak height can be determined by
\begin{eqnarray}
-\lambda_{\rm max} \eta_*
    = -\frac{\sqrt{2} B}{M_{\rm pl}} \eta_*
    = -2\sqrt{3} \frac{ B/\sqrt{2}}{\sqrt{3}M_{\rm pl}H}
   \frac{\eta_*}{\eta_1} 
   = 2\sqrt{3} \sqrt{\frac{\rho_{B*}}{\rho_{\rm inf}}} \ ,
\end{eqnarray}
where we used $a(\eta_1)=1/(H\eta_1)=1$ in the second equality. In the third equality, the energy density of the dark magnetic field at $\eta_*$ and that of the inflaton field at $\eta_1$ are written by $\rho_{B*}=(a_*/a_1)^2\,|{\bm B}|^2/2$ and $\rho_{\rm inf}=3M_{\rm pl}^2H^2$ respectively. Thus, the extra factor $\exp[-\lambda\,\eta_*]/2$ could be $16$ for $\rho_{B*}\simeq \rho_{\rm inf}$. We see that the peak height also depends on the direction of observation. The dependence on the direction of observation appears in the presence of the primordial dark magnetic fields. Hence, our results would offer new vistas to probe the primordial dark magnetic fields through the observations of primordial GWs.

\section{Conclusion}
In this paper, we studied the effect of primordial dark magnetic fields generated during inflation on the PGW power spectrum.
Assuming that there exist dark magnetic fields in the range from $10^{-17}$ to $10^{-9}$ Gauss with a cosmological coherence length, we showed that the graviton $-$dark photon conversion process affects the PGW power spectrum. The point is that the graviton dark photon conversion process induces tachyonic instability. 
As a consequence, a peak appeared in the spectrum 
and the peak height depends on the direction of observation.
Remarkably, it turned out that the peak frequency lies in the range from $10^{-5}$ to $10^{3}$ Hertz in the case of GUT scale inflation. Thus, we have a possibility to probe the primordial dark magnetic fields by observing the PGW spectrum. 

In~\cite{Watanabe:2009ct,Watanabe:2010fh,Dulaney:2010sq,Gumrukcuoglu:2010yc}, the statistical anisotropy in the PGW power spectrum is calculated in the presence of background electric fields. 
There, no peak appeared in the spectrum since there is no $k$-dependence of the coupling term in the equation of motion corresponding to Eq.~(\ref{eom:II}).
It would be interesting to study what happens when both electric and magnetic fields coexist in the background during inflation.

In this work, we have focused on the 
coupling parameter 
$c=-1/2$. In the case of $c=-1/2$, the physical dark magnetic fields decay slowly as $fB\propto 1/a$ during inflation compared to the case of constant $f$ where the physical magnetic fields decay as $fB\propto 1/a^2$. On the other hand, in the case of $c=-1$, the physical dark magnetic fields $fB$ do not decay during inflation. Hence, we expect a different shape of the PGW power spectrum.
We leave the analysis of this issue for future work.

\section*{Acknowledgments}
S.\ K. was supported by the Japan Society for the Promotion of Science (JSPS) KAKENHI Grant Number JP22K03621.
J.\ S. was in part supported by JSPS KAKENHI Grant Numbers JP17H02894, JP17K18778, JP20H01902, JP22H01220.
K.\ U. was supported by the JSPS KAKENHI Grant Number 20J22946.

\printbibliography
\end{document}